\documentclass[aps,prd,superscriptaddress,twoside,twocolumn,nofootinbib,showpacs,floatfix]{revtex4-1}
\usepackage{amsmath,amssymb}
\usepackage{graphicx}
\usepackage{subfigure}
\usepackage[colorlinks]{hyperref}
\usepackage[usenames,dvipsnames]{color}
\pdfpagewidth=\paperwidth
\pdfpageheight=\paperheight
\allowdisplaybreaks

\begin{document}

\title{\boldmath Photoproduction $\gamma p \to K^{*+} \Lambda$ in a Reggeized model}

\author{Ai-Chao Wang}
\affiliation{School of Nuclear Science and Technology, University of Chinese Academy of Sciences, Beijing 100049, China}

\author{Fei Huang}
\email[Corresponding author. Email: ]{huangfei@ucas.ac.cn}
\affiliation{School of Nuclear Science and Technology, University of Chinese Academy of Sciences, Beijing 100049, China}

\author{Wen-Ling Wang}
\affiliation{School of Physics, Beihang University, Beijing 100191, China}

\author{Guang-Xiong Peng}
\affiliation{School of Nuclear Science and Technology, University of Chinese Academy of Sciences, Beijing 100049, China}

\date{\today}

\begin{abstract}
The high-precision differential cross-section data for the reaction $\gamma p \to K^{*+}\Lambda$ are reanalyzed within a Regge-inspired effective Lagrangian approach. The model adopts Regge phenomenology to constrain the $t$-channel contributions from the $\kappa$, $K$, and $K^*$ exchanges. A minimal number of resonances in the $s$ channel are introduced in constructing the reaction amplitudes in order to describe the data. It is shown that the differential cross-section data for $\gamma p \to K^{*+}\Lambda$ can be satisfactorily described by introducing the only $N(2060){5/2}^-$ resonance in the $s$ channel, which is quite different from our earlier work performed in an effective Lagrangian approach [A. C. Wang {\it et al.}, Phys. Rev. C 96, 035206 (2017)], where the amplitudes are computed by evaluating Feynman diagrams and it is found that introducing at least one additional resonance apart from the $N(2060){5/2}^-$ is indispensable for reproducing the data. The roles of individual contributions from meson and baryon exchanges on the angular distributions are found to be highly model dependent. The extracted mass of $N(2060){5/2}^-$ turns out to be well determined, independent of how the $t$-channel amplitudes are constructed, whereas the width does not.
\end{abstract}



\maketitle

\section{Introduction}   \label{Sec:intro}

Photoproduction reaction of $\gamma p \to K^{*+} \Lambda$ provides an alternative tool to investigate excited nucleon states ($N^*$'s) besides $\pi N$ scattering and pion photoproduction. As the $K^{*+}\Lambda$ couples to $N^*$'s with isospin $1/2$ only, this reaction acts as an ``isospin filter" and is more selective to distinguish certain resonances than pion production reactions. Also, since the $K^{*+}\Lambda$ has a much higher threshold energy, this reaction is more suitable  to study the $N^*$'s in a less-explored higher mass region than pion production reactions.

Experimentally, all the available data for $\gamma p \to K^{*+}\Lambda$ have been published by the CLAS Collaboration at the Thomas Jefferson National Accelerator Facility (JLab). The preliminary cross section data have been reported in 2006 \cite{Guo:2006} and 2011 \cite{Hicks:2011}. The first high precision differential cross-section data in the energy range from threshold up to center-of-mass energy $W=2.85$ GeV were reported in 2013 \cite{Tang:2013}.

Theoretically, the $K^{*+}$ photoproduction on the proton was first studied in a quark model in 2001 \cite{Zhao:2001}. In Refs.~\cite{Oh-1:2006,Oh-2:2006}, the preliminary total cross section data for the $\gamma p \to K^{*+} \Lambda$ reported by the CLAS Collaboration in 2006 \cite{Guo:2006} were analyzed within an isobar model. Later in 2010, a Regge model without considering any nucleon resonances was proposed to study the preliminary total cross section data, and it is claimed that the $K^*$ trajectory exchange and the contact term provide sizable contributions \cite{Ozaki:2010}. In Ref.~\cite{Kim:2011}, the preliminary differential cross-section data reported by the CLAS Collaboration in 2011 \cite{Hicks:2011} were analyzed within an effective Lagrangian approach. When the first high precision differential cross-section data were reported in 2013 \cite{Tang:2013}, it was found that all the previously published theoretical works underestimated the cross sections in the range of photon laboratory energy $2.1$ $<E_\gamma<$ $3.1$ GeV. In Ref.~\cite{Kim:2014}, Kim {\it et al.} reinvestigated the $\gamma p \to K^{*+}\Lambda$ reaction by a fit to the high precision differential cross-section data with several nucleon resonances considered in an effective Lagrangian approach. The so far best theoretical description of the high precision differential cross-section data for the $\gamma p \to K^{*+}\Lambda$ was published by our group in Ref.~\cite{Wang:2017}, where a detailed review of the status of experimental and theoretical studies of the $\gamma p \to K^{*+} \Lambda$ reaction can also be found. In Ref.~\cite{Yu:2017}, a Regge model is proposed for this reaction focusing on the forward angle behavior of the angular distributions and thus no $N^*$'s have been introduced in constructing the reaction amplitudes. Note that in the Regge model of Refs.~\cite{Ozaki:2010,Yu:2017}, since no nucleon resonances are considered, the details of the angular distributions, especially the near-threshold structures of the differential cross sections, are not expected to be well described. In Ref.~\cite{Hejun:2016}, the $K^{*0}$ photoproduction off the neutron was studied.

In the literature, models based on the meson-exchange picture have been widely used in investigating the pseudoscalar meson and vector meson photoproduction reactions. In these models, a reliable extraction of resonance contents and the associated resonance parameters suffers from the fact that the background contributions, which play a substantial role in the reaction mechanism, cannot be constrained in a model-independent way. Of course, as is well known, the $t$-channel or $u$-channel amplitudes are usually quite distinctive if the considered meson or baryon exchanges are different in various models. Nevertheless, there is a case in which, even if the exchanged mesons or baryons are the same, the $t$-channel or $u$-channel amplitudes may still be quite different, because the meson or baryon exchange amplitudes can be constructed in totally different ways. In most cases, the $t$-channel meson exchanges or the $u$-channel baryon exchanges are described by Feynman amplitudes in a tree-level approximation \cite{Kim:2014,Wang:2017,Wang:2018}. The so-called Regge amplitudes, where the Regge propagators are introduced to replace the traditional Feynman propagators, are also commonly used to describe the $t$-channel or $u$-channel interactions even in the low-energy regions \cite{Chiang:2003,Ozaki:2010,Yu:2017}. In recent years, a hybrid approach called the interpolated Regge model was applied to reproduce the data for the $\Lambda(1520)$ and $\Sigma(1385)$ photoproductions \cite{Toki:2008,Nam:2010,He:2012,He:2014}, where in the $t$ channel an interpolating form factor was introduced so that the amplitudes behave as Feynman amplitudes at low energies and Regge amplitudes at high energies. In literature, these three different types of $t$-channel amplitudes, i.e., the Feynman amplitudes, the Regge amplitudes and the interpolated Regge amplitudes, are employed by different groups to describe data for different reactions. Nevertheless, it is so far not clear how different the results will be for a specific reaction if one construct the $t$-channel amplitudes in different ways. In particular, one is interested in how strongly the reaction mechanism, the resonance contents and the associated resonance parameters depend on the particular way to construct the $t$-channel amplitudes.

In Ref.~\cite{Wang:2017}, we have presented a good description of the high-precision differential cross-section data for the $\gamma p \to K^{*+}\Lambda$ within an effective Lagrangian approach. One of the most prominent features of this work is that the near-threshold structures exhibited by the differential cross-section data are satisfactorily described for the first time. It is found that it is the $t$-channel $K$ exchange and the $s$-channel $N(2060)5/2^-$ exchange that dominate the dynamics of the $\gamma p \to K^{*+}\Lambda$ reaction in the near-threshold region. Moreover, apart from the $N(2060)5/2^-$ resonance, an additional resonance, which could be one of the $N(2000)5/2^+$, $N(2040)3/2^+$, $N(2100)1/2^+$, $N(2120)3/2^-$, and $N(2190)7/2^-$ resonances, is also found to be indispensable to describe the data.

The $t$-channel meson exchanges in Ref.~\cite{Wang:2017} are described by the Feynman amplitudes. As mentioned above, it is worth investigating whether the reaction mechanism keeps the same and, in particular, how much the extracted resonance contents and parameters change if the $t$-channel meson exchanges are described by the Regge amplitudes or the interpolated Regge amplitudes instead of the Feynman amplitudes.

In the present work, we reanalyze the high-precision differential cross-section data for $\gamma p \to K^{*+} \Lambda$ within a Regge-inspired effective Lagrangian approach. Both the traditional Regge amplitudes and the so-called interpolated Regge amplitudes are employed for the $t$-channel $\kappa$, $K$, and $K^*$ exchanges. The purpose is to test how strongly the reaction mechanism, the extracted resonance contents and the associated resonance parameters depend on the way of constructing the $t$-channel meson-exchange amplitudes.

We mention that a coupled-channel approach is required for a more complete analysis of the data and a more reliable extraction of the properties of nucleon resonances. So far, such an approach has been developed mostly for pseudoscalar meson production reactions \cite{Anisovich:2012,Ronchen:2013,Shklyar:2013,Kamano:2015,Ramirez:2016}. The tree-level calculation performed in the present work may be considered as a starting point toward developing such a more complete model.

The paper is organized as follows. In the next section, we briefly introduce the formalism of the traditional Regge model and the interpolated Regge model. The numerical results of the cross sections from these two models are presented in Sec.~\ref{Sec:results}, where a comparison of the results from these two models and those from the model of Ref.~\cite{Wang:2017} is also given. Finally, a summary and conclusions are given in Sec.~\ref{sec:summary}.

\section{Formalism}  \label{Sec:formalism}

\begin{figure}[tbp]
\subfigure[~$s$ channel]{\includegraphics[width=0.45\columnwidth]{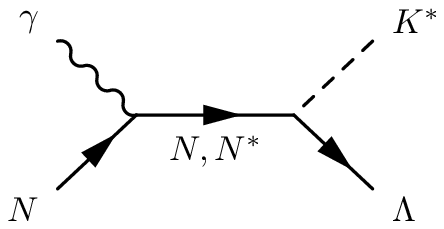}}  {\hglue 0.4cm}
\subfigure[~$t$ channel]{\includegraphics[width=0.45\columnwidth]{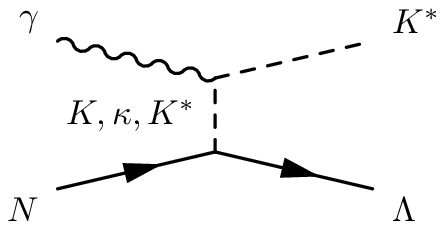}} \\[6pt]
\subfigure[~$u$ channel]{\includegraphics[width=0.45\columnwidth]{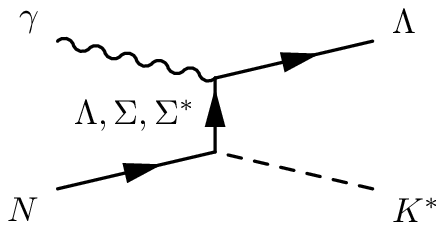}} {\hglue 0.4cm}
\subfigure[~Interaction current]{\includegraphics[width=0.45\columnwidth]{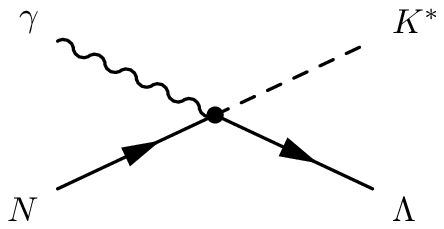}}
\caption{Generic structure of the amplitude for $\gamma N\to K^{*}\Lambda$. Time proceeds from left to right.}
\label{FIG:feymans}
\end{figure}

Following our previous work \cite{Wang:2017}, we consider the following contributions in constructing the $s$-, $t$-, and $u$-channel amplitudes for the $\gamma N\to K^*\Lambda$ reaction: (a) the $N$ and $N^*$'s exchanges in the $s$ channel, (b) the $K$, $\kappa$, and $K^*$ meson exchanges in the $t$ channel, and (c) the $\Lambda$, $\Sigma$, and $\Sigma^*(1385)$ hyperon exchanges in the $u$ channel. These contributions are schematically shown in Fig.~\ref{FIG:feymans}. The full reaction amplitude can then be expressed as
\begin{eqnarray}
M^{\nu\mu} = M^{\nu\mu}_s + M^{\nu\mu}_t + M^{\nu\mu}_u + M^{\nu\mu}_{\rm int},  \label{eq:amplitudes}
\end{eqnarray}
with $M^{\nu\mu}_s$, $M^{\nu\mu}_t$, and $M^{\nu\mu}_u$ standing for $s$-, $t$-, and $u$-channel amplitudes, respectively, and $M^{\nu\mu}_{\rm int}$ representing the generalized interaction current which is introduced to keep the full amplitude $M^{\nu\mu}$ gauge invariant \cite{Haberzettl:1997,Haberzettl:2006,Huang:2012,Huang:2013}. The explicit expression of $M^{\nu\mu}_{\rm int}$ and the effective Lagrangians, propagators and form factors introduced to evaluate the Feynman diagrams of $M^{\nu\mu}_s$, $M^{\nu\mu}_t$, and $M^{\nu\mu}_u$ can be found in Ref.~\cite{Wang:2017}. Here we just present the new formalism that is relevant to the Reggeized treatment of the $t$-channel $K$, $K^*$, and $\kappa$ exchanges.

\subsection{Reggeized $t$-channel amplitudes} \label{subsec:Regge}

The most economic way to take into account the effects of high spin meson exchanges is to substitute the $t$-channel Feynman amplitudes by Regge amplitudes. The standard Reggeization of the $t$-channel $K$, $\kappa$, and $K^*$ exchanges corresponds to the following replacements:
\begin{align}
\frac{1}{t-m^2_K}  ~\Longrightarrow~  {\cal P}^K_{\rm Regge} = & \left(\frac{s}{s_0}\right)^{\alpha_K(t)} \frac{\pi \alpha'_K}{\sin[\pi\alpha_K(t)]}  \nonumber \\
& \times \frac{1}{\Gamma[1+\alpha_K(t)]},   \label{eq:Regge_prop_K} \\[6pt]
\frac{1}{t-m^2_\kappa}  ~\Longrightarrow~ {\cal P}^\kappa_{\rm Regge} = & \left(\frac{s}{s_0}\right)^{\alpha_\kappa(t)} \frac{1 + e^{-i\pi\alpha_\kappa(t)}}{2} \nonumber \\
& \times  \frac{\pi \alpha'_\kappa}{\sin[\pi\alpha_\kappa(t)]} \frac{1}{\Gamma[1+\alpha_\kappa(t)]}, \label{eq:Regge_prop_kappa}  \\[6pt]
\frac{1}{t-m^2_{K^*}}  ~\Longrightarrow~ {\cal P}^{K^*}_{\rm Regge} = & \left(\frac{s}{s_0}\right)^{\alpha_{K^*}(t)-1} \frac{1 - e^{-i\pi\alpha_{K^*}(t)}}{2} \nonumber \\
& \times  \frac{\pi \alpha'_{K^*}}{\sin[\pi\alpha_{K^*}(t)]} \frac{1}{\Gamma[\alpha_{K^*}(t)]}.   \label{eq:Regge_prop_Kstar}
\end{align}
Here $s_0$ is a mass scale which is conventionally taken as $s_0=1$ GeV$^2$, and $\alpha'_M$ is the slope of the Regge trajectory $\alpha_M(t)$. For $M=K$, $\kappa$, and $K^*$, the trajectories are parametrized as \cite{Corthals:2007}
\begin{align}
\alpha_K(t) &= 0.7~{\rm GeV}^{-2}\left(t-m_K^2\right),   \label{eq:trajectory_K}  \\[3pt]
\alpha_\kappa(t) &= 0.7~{\rm GeV}^{-2} \left(t-m_\kappa^2\right),  \label{eq:trajectory_kappa}  \\[3pt]
\alpha_{K^*}(t) &= 1 + 0.85~{\rm GeV}^{-2} \left(t-m_{K^*}^2\right).  \label{eq:trajectory_Kstar}
\end{align}
Note that in Eq.~(\ref{eq:Regge_prop_K}) a degenerate trajectory is employed for the $K$ exchange, thus the signature factor reduces to $1$. Such a choice is preferred by data, which has been tested by our numerical calculation.

In Ref.~\cite{Wang:2017}, an appropriate prescription for the interaction current $M^{\nu\mu}_{\rm int}$ is chosen to guarantee that the full photoproduction amplitude $M^{\nu\mu}$ for $\gamma p \to K^{*+}\Lambda$ satisfies the generalized Ward-Takahashi identity and thus is fully gauge invariant \cite{Haberzettl:1997,Haberzettl:2006,Huang:2012,Huang:2013}. Note that our prescription for $M^{\nu\mu}_{\rm int}$ is independent of any particular form of the $t$-channel form factor, $f_t$, provided that it is normalized as $f_t\left(t=m_{K^*}^2\right)=1$. One observes that the Reggeization of the amplitude for $K^*$ exchange in Eq.~(\ref{eq:Regge_prop_Kstar}) is equivalent to the following replacement:
\begin{equation}
f_t ~\Longrightarrow~ {\cal F}_t = \left(t-m_{K^*}^2\right) {\cal P}^{K^*}_{\rm Regge}(t) f_t.  \label{eq:form_factor}
\end{equation}
Therefore, simply replacing $f_t$ by ${\cal F}_t$ in the prescription of $M^{\nu\mu}_{\rm int}$ in Ref.~\cite{Wang:2017}, the gauge invariance requirement will still be satisfied for the Reggeized amplitude \cite{Haberzettl:2015}. Explicitly we have
\begin{equation}
M^{\nu\mu}_{\rm int} = \Gamma^\nu_{\Lambda N K^*}(q) C^\mu + M_{\rm KR}^{\nu\mu} {\cal F}_t,  \label{eq:Mint_r}
\end{equation}
where the auxiliary current $C^\mu$ reads
\begin{equation}
C^\mu =  - Q_{K^*} \frac{{\cal F}_t-\hat{F}}{t-q^2}  (2q-k)^\mu - Q_N \frac{f_s-\hat{F}}{s-p^2} (2p+k)^\mu, \label{eq:Mint_r2}
\end{equation}
and $\hat{F}$ is
\begin{equation} \label{eq:Fhat}
\hat{F} = 1 - \hat{h} \left(1 -  f_s\right) \left(1 - {\cal F}_t\right).
\end{equation}
As usual, the parameter $\hat{h}$ is chosen to be $1$ for simplicity. $f_s$ is the form factor for the $s$-channel nucleon exchange, $\Gamma^\nu_{\Lambda N K^*}(q)$ is the $\Lambda N K^*$ vertex with $q$ being the momentum for the outgoing $K^*$, and $M_{\rm KR}^{\nu\mu}$ is the Kroll-Ruderman term. We refer the readers to Ref.~\cite{Wang:2017} for more details.

\subsection{Interpolated $t$-channel Regge amplitudes} \label{subsec:interpolated_Regge}

Considering that the Regge amplitudes work properly in the very-large-$s$ and very-small-$|t|$ region, while the Feynman amplitudes work properly in the low energy region, in Refs.~\cite{Toki:2008,Nam:2010,He:2012,He:2014}, an interpolating form factor is introduced to parametrize the smooth transition from the Feynman amplitudes to the Regge amplitudes. Instead of Eq.~(\ref{eq:form_factor}), one has the following replacement for the Feynman amplitudes:
\begin{equation}
f_t ~\Longrightarrow~ {\cal F}_{R,t} = {\cal F}_t R + f_t \left(1-R\right),   \label{Eq: intRe}
\end{equation}
where $R=R_sR_t$ with
\begin{align}
R_s &= \frac{1}{1+e^{-\left(s-s_R\right)/s_0}},  \\[3pt]
R_t &= \frac{1}{1+e^{-\left(t+t_R\right)/t_0}}.
\end{align}
Here $s_R$, $s_0$, $t_R$, and $t_0$ are parameters to be fixed by fitting the data.

It is easy to see that by making the replacement in Eq.~(\ref{Eq: intRe}), the $t$-channel amplitudes will be a combination of the Regge amplitudes and the Feynman amplitudes, with a weight $R$ for the former and $(1-R)$ for the latter. In the low-energy and small-$|t|$ region, the factor $f_t R$ tends towards $0$, ensuring that one has almost pure Feynman amplitudes. In the high-energy and small-$|t|$ region, the factor $f_t (1-R)$ tends towards $0$, ensuring that one has almost pure Regge amplitudes. In the intermediate energy region, the amplitudes are calculated as a mixture of the Feynman amplitudes and the Regge amplitudes, which could be considered as a phenomenological prescription for the physical amplitudes in this energy region and it depends on the newly introduced parameters $s_R$, $s_0$, $t_R$, and $t_0$.

As discussed in Sec.~\ref{subsec:Regge}, the gauge invariance for the full amplitude is still kept when the replacement of Eq.~(\ref{Eq: intRe}) is also performed in the interaction current $M^{\nu\mu}_{\rm int}$.

\section{Results and discussion}   \label{Sec:results}

\begin{figure*}[tbp]
\includegraphics[width=0.65\textwidth]{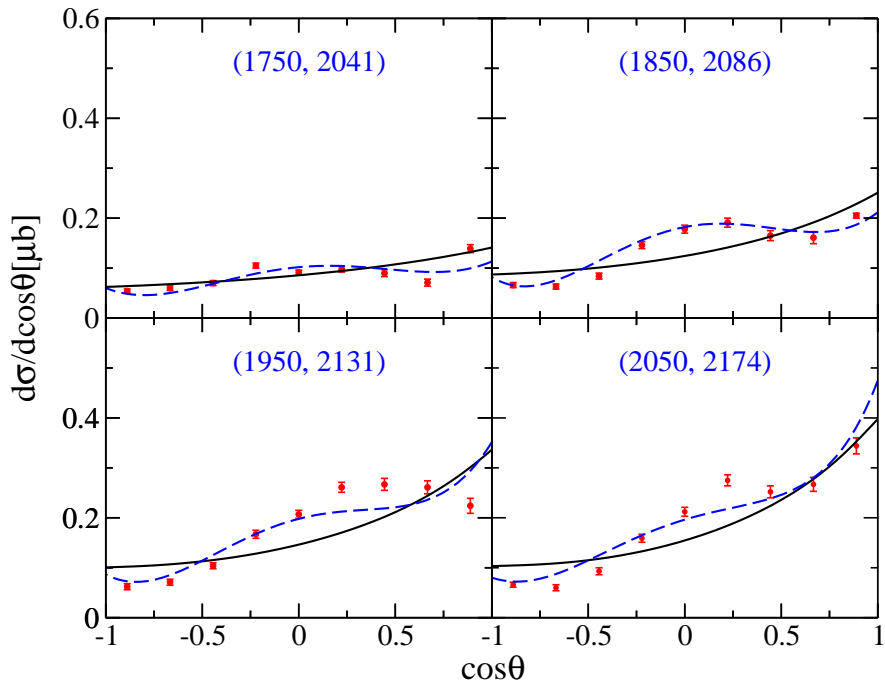}
\caption{Differential cross sections for $\gamma p \to K^{*+}\Lambda$ as a function of $\cos\theta$ in the center-of-mass frame in the near-threshold region. The black solid lines and the blue dashed lines correspond to the  results fitted with the $N(2100)1/2^+$ resonance and the results fitted with the $N(2060)5/2^-$ resonance in model II (the interpolated Regge model), respectively. The scattered symbols denote the CLAS data from Ref.~\cite{Tang:2013}. The numbers in parentheses denote the photon laboratory incident energy (left number) and the total center-of-mass energy of the system (right number), in MeV.}
\label{fig:fig_1p}
\end{figure*}

\begin{table*}[tb]
\caption{\label{Table:para1} Fitted values of the parameters in model I (Regge model) and model II (interpolated Regge model). For comparison, the corresponding values of the parameters in model I of Ref.~\cite{Wang:2017} (Feynman model) are also listed in the last column. Here $\beta_{\Lambda K^*}$ is the branching ratio for the decay $N(2060)5/2^-\to\Lambda K^*$. $A_{1/2}$ and $A_{3/2}$ are the helicity amplitudes for the radiative decay $N(2060)5/2^-\to\gamma p$. Note that in model I of Ref.~\cite{Wang:2017}, $\Lambda_{K^*}$ and $\Lambda_\kappa$ are fixed to be $900$ and $1100$ MeV, respectively.  }
\renewcommand{\arraystretch}{1.2}
\begin{tabular*}{\textwidth}{@{\extracolsep\fill}lrrr}
\hline\hline
       &    Model I    &   Model II & Model I of Ref.~\cite{Wang:2017} \\
\hline
$g^{(1)}_{\Sigma^* \Lambda \gamma}$  & $-2.28\pm 0.07$ & $-2.09\pm 0.09$ & $0.74\pm 0.16$  \\
$\Lambda_{K^*}$ [MeV]   & $900\pm 24$    &   $752\pm 2$ &   ${\bf 900}$  \\
$\Lambda_K$ [MeV]   & $1940\pm 25$    &   $762\pm 12$ &   $1000\pm 6$   \\
$\Lambda_{\kappa}$ [MeV]   & $900\pm 64$    &   &   ${\bf 1100}$ \\
\hline
$N(2060){5/2}^-$ parameters  &     &      \\[-3pt]
$M_R$ [MeV]              & $2033\pm 4$     & $2028\pm 2$ & $2033\pm 2$     \\
$\Gamma_R$ [MeV]   & $124\pm 7$         &  $57\pm 2$ &  $65\pm4$         \\
$\Lambda_R$ [MeV]   & $1270\pm 6$   &  $1200\pm 5$ &  $1188\pm 20$   \\
$\sqrt{\beta_{\Lambda K^*}}A_{1/2}$ [$10^{-3}$\,GeV$^{-1/2}$]  & $0.66\pm 0.09$  & $0.84\pm 0.08$ & $0.69\pm 0.06$ \\
$\sqrt{\beta_{\Lambda K^*}}A_{3/2}$ [$10^{-3}$\,GeV$^{-1/2}$]  & $-0.93\pm 0.14$  &  $-1.19\pm 0.11$ & $-1.39\pm 0.13$  \\
\hline\hline
\end{tabular*}
\end{table*}

\begin{table}[tb]
\caption{\label{Table:para2} Fitted values of the parameters in interpolated form factor in model II (interpolated Regge model).}
\renewcommand{\arraystretch}{1.2}
\begin{tabular*}{\columnwidth}{@{\extracolsep\fill}cccc}
\hline\hline
$s_0$ [GeV$^2$] & $s_R$ [GeV$^2$] & $t_0$ [GeV$^2$] & $t_R$ [GeV$^2$] \\
$0.66\pm 0.06$ & $4.95\pm 0.07$ & $0.25\pm 0.01$ & $0.46\pm 0.02$ \\
\hline\hline
\end{tabular*}
\end{table}

\begin{figure*}[tbp]
\includegraphics[width=0.85\textwidth]{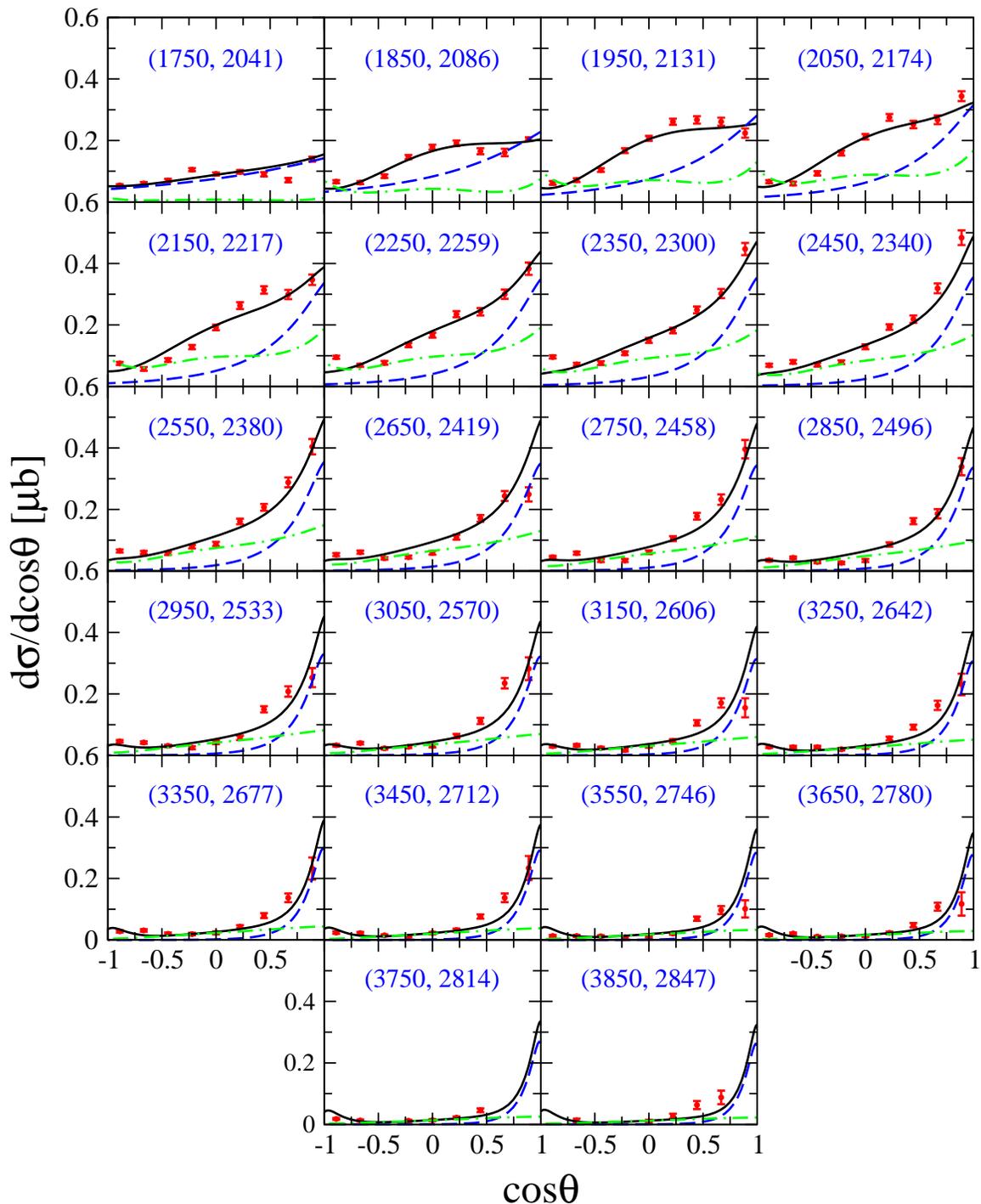}
\caption{Differential cross sections for $\gamma p \to K^{*+}\Lambda$ as a function of $\cos\theta$ (black solid lines) from model I (Regge model). The scattered symbols denote the CLAS data from Ref.~\cite{Tang:2013}. The blue dashed and green dash-dotted lines represent the individual contributions from the $K$ and $N(2060)5/2^-$ exchanges, respectively. The numbers in parentheses denote the photon laboratory incident energy (left number) and the corresponding total center-of-mass energy of the system (right number), in MeV.}
\label{fig:dif1}
\end{figure*}

\begin{figure*}[tbp]
\includegraphics[width=0.85\textwidth]{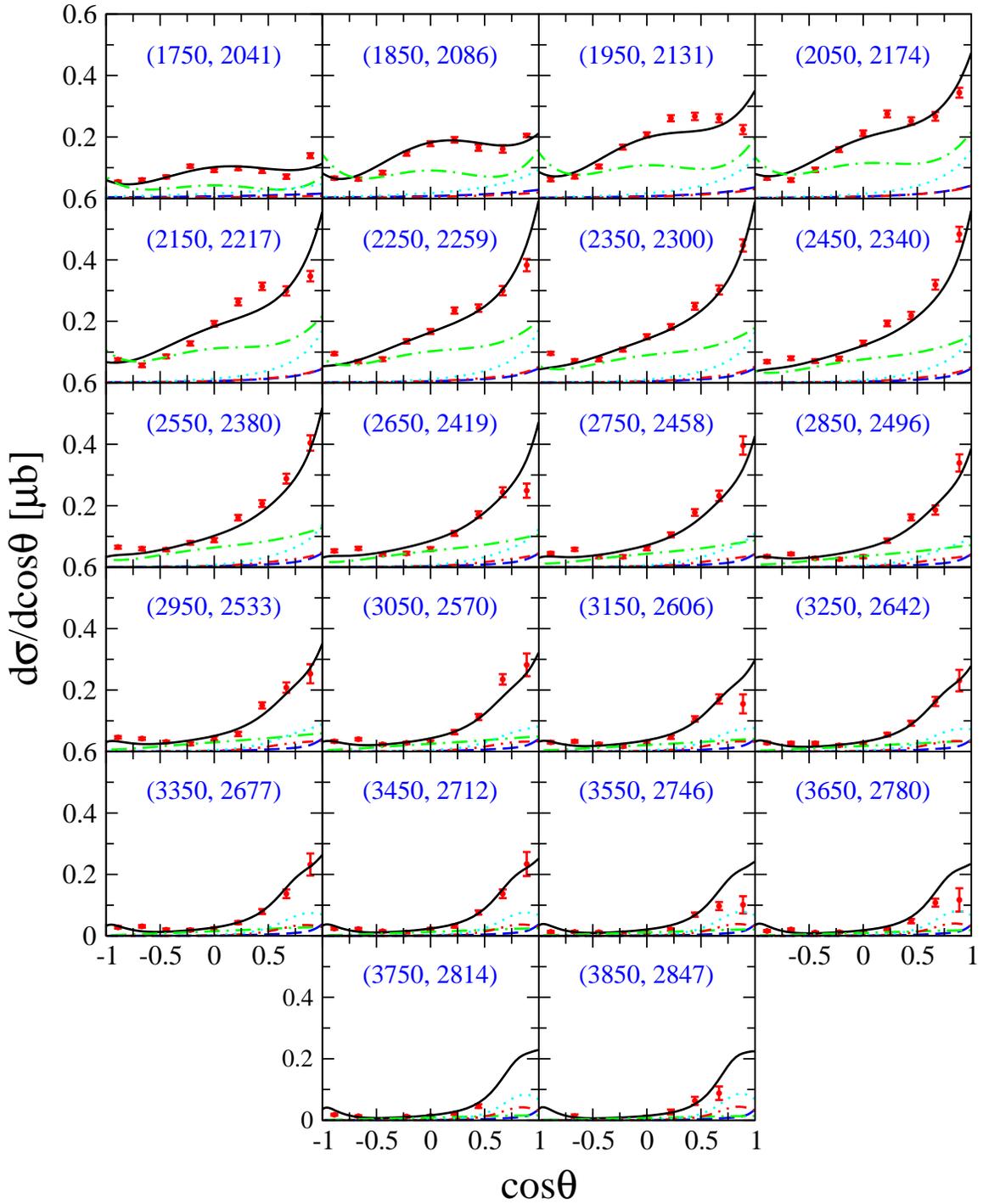}
\caption{Differential cross sections for $\gamma p \to K^{*+}\Lambda$ as a function of $\cos\theta$ (black solid lines) from model II (interpolated Regge model). In addition to the same notations as in Fig.~\ref{fig:dif1}, the red dash-double-dotted and cyan dotted lines represent the contributions from the $K^*$ exchange and the interaction current, respectively.}
\label{fig:dif2}
\end{figure*}

\begin{figure*}[tbp]
\includegraphics[width=0.95\textwidth]{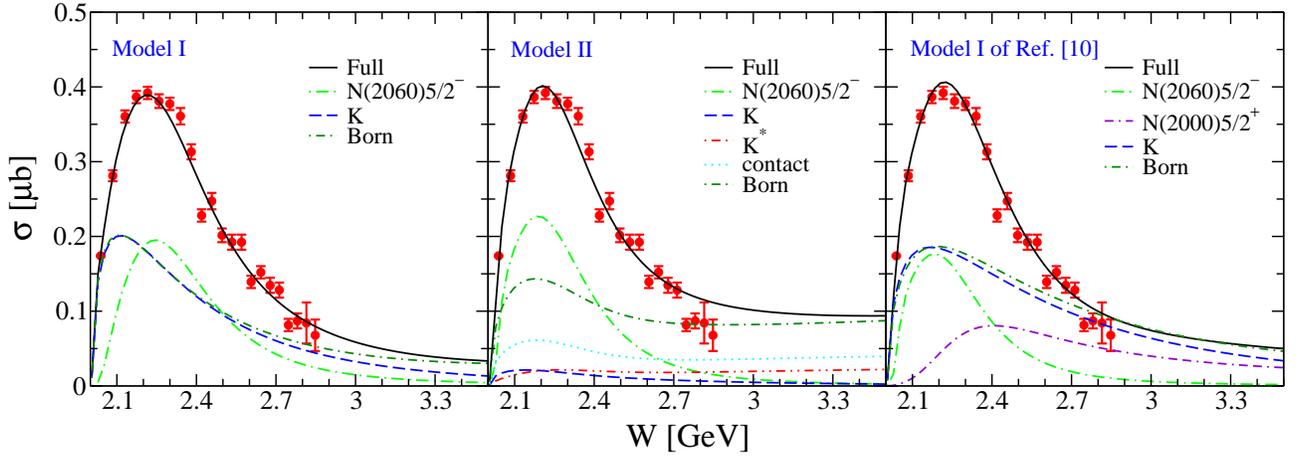}
\caption{Total cross sections with individual contributions for  $\gamma p \to K^{*+}\Lambda$. The left graph is for model I (Regge model), the middle one corresponds to model II (interpolated Regge model), and the right one is for model I of Ref.~\cite{Wang:2017} (Feynman model). The black solid lines represent the full results. The green dash-dotted, blue dashed, red dash-double-dotted, and cyan dotted lines represent the individual contributions from the $N(2060)5/2^-$, $K$, $K^*$ exchanges and the interaction current, respectively. The olive short-dash-dotted lines represent the contributions from the Born term which consists of the coherent sum of all the contributions other than the $s$-channel resonance exchanges. In the right graph, the purple double-dash-dotted line represents the contributions from the $N(2000)5/2^+$ exchange. The contributions from the $K^*$ exchange and the interaction current in the left and right graphs are not presented, and so are the contributions from other terms in these three graphs, since these contributions are too small to be clearly seen with the scale used in this figure. The scattered symbols are data from CLAS Collaboration \cite{Tang:2013}. }
\label{fig:total_cro_sec}
\end{figure*}

For conciseness of the discussions, we refer to the model discussed in Sec.~\ref{subsec:Regge} as the Regge model (model I),  and the model discussed in Sec.~\ref{subsec:interpolated_Regge} as the interpolated Regge model (model II).

In Ref.~\cite{Wang:2017}, a rather satisfactory description of the differential cross section data for the $\gamma p \to K^{*+}\Lambda$ has been achieved within an effective Lagrangian approach with the amplitudes being constructed by use of Feynman propagators. In this work, we reanalyze the high-precision differential cross-section data for this reaction within a Regge-inspired effective Lagrangian approach. The $t$-channel Feynman amplitudes in Ref.~\cite{Wang:2017} have now been replaced by either the Regge amplitudes as discussed in Sec.~\ref{subsec:Regge} or the interpolated Regge amplitudes as discussed in Sec.~\ref{subsec:interpolated_Regge}. The purpose is to investigate how strongly the reaction mechanism, the extracted resonance contents and the associated resonance parameters depend on the way of constructing the $t$-channel amplitudes.

As has been done in Ref.~\cite{Wang:2017}, we introduce a minimal number of resonances in the $s$ channel in constructing the reaction amplitudes in order to describe the data. If no nucleon resonance is taken into account, it turns out that one cannot reproduce the shape of the angular distribution near the $K^{*+}\Lambda$ threshold exhibited by the CLAS data with reasonable model parameters. We then try to consider one nucleon resonance. We test one by one the $N(2000)5/2^+$, $N(2040)3/2^+$, $N(2060)5/2^-$, $N(2100)1/2^+$, $N(2150)3/2^-$, and $N(2190)7/2^-$ resonances which are located near the $K^{*+}\Lambda$ threshold and might potentially contribute to the $\gamma p \to K^{*+}\Lambda$ reaction. After numerous tests, it is found that in both the Regge model and the interpolated Regge model, the differential cross-section data for $\gamma p \to K^{*+}\Lambda$ can be satisfactorily described by introducing the $N(2060){5/2}^-$ resonance, which is quite different from Ref.~\cite{Wang:2017}, where it was found that apart from the $N(2060){5/2}^-$ resonance, in order to reproduce the data one needs to introduce in the $s$ channel at least one additional resonance, which could be one of the $N(2000)5/2^+$, $N(2040)3/2^+$, $N(2100)1/2^+$, $N(2120)3/2^-$, and $N(2190)7/2^-$ resonances. The fits with the inclusion of other nucleon resonances are not considered acceptable, not only because they have much larger $\chi^2$, but also because in these fits the shape of the angular distributions near threshold cannot be reproduced. As an illustration, we show in Fig. \ref{fig:fig_1p} a comparison of the near-threshold differential cross sections from the fit with the $N(2060)5/2^-$ resonance and those from the fit with the $N(2100)1/2^+$ resonance in model II. One sees clearly that the fit with the $N(2100)1/2^+$ resonance fails to describe the near-threshold differential cross sections and thus is not treated as an acceptable fit.

The values of resonance parameters of the $N(2060){5/2}^-$ and other adjustable parameters determined in both the Regge model and the interpolated Regge model are listed in Tables~\ref{Table:para1} and \ref{Table:para2}. Note that the contributions from the $\kappa$ exchange are found to be negligible in the interpolated Regge model and thus they are not included in constructing the reaction amplitudes. The other relevant parameters are nonadjustable, and they are fixed by flavor SU(3) relations or taken from other references, as explained in Ref.~\cite{Wang:2017}. For comparison, the corresponding values of the parameters in model I of Ref.~\cite{Wang:2017} (the Feynman model) are also listed in the last column of Table~\ref{Table:para1}. Note that in this model, $\Lambda_{K^*}$ and $\Lambda_\kappa$ are fixed to be $900$ and $1100$ MeV, respectively. One sees that the parameters in the background contributions are quite different in model I, model II, and model I of Ref.~\cite{Wang:2017}. For example, the fitted value for $\Lambda_K$ is $1940$ MeV in model I, but it is $1000$ MeV in model I of Ref.~\cite{Wang:2017}. This does not mean that the contributions from the $K$ exchange in model I are much bigger than those in model I of Ref.~\cite{Wang:2017}, since the corresponding amplitudes in these two models are constructed in quite different ways, i.e., one is constructed in the Regge type while the other is constructed in the Feynman type. Actually, as will be discussed in connection with the total cross section (cf. Fig.~\ref{fig:total_cro_sec}), the contributions from the $K$ exchange in these two models are comparable, although they are not the same. The values of the coupling constant $g_{\Sigma^*\Lambda\gamma}^{(1)}$ in model I and model II are quite different from that in model I of Ref.~\cite{Wang:2017}. This is mainly because the $u$-channel $\Sigma^*$ exchange has rather tiny contributions and thus they are not well constrained by the available cross-section data in the considered reaction. Note that the coupling constant $g_{\Sigma^*\Lambda\gamma}^{(2)}$ is constrained by the $\Sigma^{*0}\to \Lambda\gamma$ decay width, $\Gamma_{\Sigma^{*0}\to \Lambda\gamma}=0.45$ MeV, which results in $g_{\Sigma^*\Lambda\gamma}^{(2)}=-22.6$, $-25.23$, and $40.76$ in model I, model II, and model I of Ref.~\cite{Wang:2017}, respectively. For the $N(2060){5/2}^-$ resonance, the masses determined in these three models are consistent with each other, indicating that this parameter is well determined, independently of how the $t$-channel amplitudes are constructed. The values of the cutoff parameter and the reduced helicity amplitudes for this resonance are also found to be close to each other in these three models. The values of the $N(2060){5/2}^-$ width vary in different models, indicating that this parameter cannot be well determined by the considered data.

The differential cross sections produced in both the Regge model and the interpolated Regge model are shown in Figs.~\ref{fig:dif1} and \ref{fig:dif2}, respectively, where the major individual contributions are also shown. In these two figures, the black solid lines represent the full results. The blue dashed and the green dash-dotted lines show the contributions from the $K$ and $N(2060){5/2}^-$ exchanges, respectively. In Fig.~\ref{fig:dif2}, the red dash-double-dotted and the cyan dotted lines represent the contributions from the $K^*$ exchange and the interaction current, respectively.

From Figs.~\ref{fig:dif1} and \ref{fig:dif2}, one sees that the overall fitting qualities of both the Regge model and the interpolated Regge model are satisfactory: both are in agreement with the data and comparable with that from the Feynman model of Ref.~\cite{Wang:2017}. Nevertheless, at low energies near the $\Lambda K^*$ threshold, especially at the center-of-mass energy $W=2041$ MeV, the results from the interpolated Regge model are much closer to the data than those from the Regge model and the Feynman model of Ref.~\cite{Wang:2017}.

The reaction mechanisms from the Regge model, the interpolated Regge model and the Feynman model of Ref.~\cite{Wang:2017} are quite different, as can be seen from the dominant individual contributions of the differential cross sections. In the near-threshold region, the angular distributions are dominated by the $K$ exchange in the Regge model and $N(2060){5/2}^-$ exchange in the interpolated Regge model, while in the Feynman model of Ref.~\cite{Wang:2017} both the $K$ and $N(2060){5/2}^-$ exchanges are important in the near-threshold region. At higher energies, in the Regge model the $K$ exchange still dominates the reaction, and the $N(2060){5/2}^-$ exchange also offers significant contributions with a maximum located around $2.2$--$2.3$ GeV. The contributions other than the $K$ and $N(2060){5/2}^-$ exchanges are tiny in this model. In the interpolated Regge model, the contributions from the $K$ exchange are rather small, and the dominant contributions are from the $N(2060){5/2}^-$ exchange in the energy range from threshold up to $W\approx 2.6$ GeV. The $K^*$ exchange and the interaction current also provide considerable contributions in this model in almost the whole energy region considered. In the Feynman model of Ref.~\cite{Wang:2017}, at higher energies, the $K$ exchange dominates the reaction, while the $N(2060){5/2}^-$ and the other resonance exchanges also provide considerable contributions. The $\kappa$, $K^*$ exchanges and the interaction current offer rather small contributions in the Feynman model, similarly to the Regge model.

The $N(2060){5/2}^-$ is the common resonance required in all the Regge model, the interpolated Regge model and the Feynman model of Ref.~\cite{Wang:2017}. In the last model, the fitted mass of the $N(2060){5/2}^-$ is around $2009$--$2043$ MeV and the fitted width is around $65$--$213$ MeV in different fits. From Table~\ref{Table:para1} one sees that the fitted mass of this resonance is $2033$ MeV in the Regge model and $2028$ MeV in the interpolated Regge model. An interesting observation is that these two values are very close to each other and they are both in the range predicted by the Feynman model of Ref.~\cite{Wang:2017}. The fitted width of the $N(2060){5/2}^-$ is $124$ MeV in the Regge model and $57$ MeV in the interpolated Regge model. Although not close to each other, they both are almost in the range predicted by the Feynman model of Ref.~\cite{Wang:2017}. This finding might be a hint that the $N(2060){5/2}^-$ is really needed and plays a significant role in the reaction $\gamma p\to K^{*+}\Lambda$, independently of how the $t$-channel amplitudes are constructed. The mass of this resonance can be basically determined by the present data for this reaction, while its width cannot be well determined with the present data.

Figure~\ref{fig:total_cro_sec} shows the predicted total cross sections (black solid lines) together with the individual contributions from the $K$ exchange (blue dashed lines), the $N(2060){5/2}^-$ exchange (green dash-dotted lines), the $K^*$ exchange (the red dash-double-dotted lines), the interaction current (cyan dotted lines) and the Born term (olive short-dash-dotted lines) obtained by integrating the corresponding differential cross sections in both the Regge model (model I) and the interpolated Regge model (model II). The Born terms consist of the coherent sum of all the contributions other than the $s$-channel resonance exchanges. For comparison, we also show in this figure the major contributions from model I of Ref.~\cite{Wang:2017}, illustrating the results from a Feynman model. In this model, the contributions from the $N(2000)5/2^+$ exchange are plotted with the double-dash-dotted line. The contributions from the $K^*$ exchange and the interaction current in the left and right graphs are not presented, and so are the contributions from the other terms in these three graphs, since these contributions are too small to be clearly seen with the scale used in this figure. To see the high-energy behavior of the theoretical total cross sections, we extend the plot up to $W=3.5$ GeV. One sees that in all the three models, the predicted total cross sections are in fairly good agreement with the data over the entire energy region considered, except that around $W\approx 2.8$ GeV where the theoretical total cross sections from model II overestimates the corresponding data. Note that the CLAS total cross section data are obtained by integrating the measured differential cross sections, and thus they may suffer from the limited angular acceptance of the CLAS detector at forward angles \cite{Tang:2013}. In the energy region $W>2.7$ GeV, model II produces much bigger total cross sections than the other two models. More elaborate data  in this energy region may help distinguish the theoretical models. The contributions from the Born terms are found to be important in all the three models, and in particular, they dominate the total cross sections at high energies. Nevertheless, the Born terms have different origins in various models. The $K$ exchange is seen to play a dominant role in the Regge model (model I) and the Feynman model (model I of Ref.~\cite{Wang:2017}), while it provides rather small contributions in the interpolated Regge model (model II). It also shows that the contributions from the $K$ exchange drop much faster in the Regge model than those in the Feynman model with the increase of the center-of-mass energy. In the interpolated Regge model (model II), considerable contributions are also seen from the $K^*$ exchange and the interaction current. These two together with the $K$ exchange are the major sources of the Born term in this model. In all the three models, the $N(2060){5/2}^-$ exchange offers rather important contributions, but the locations and the strengths of the bumps caused by this resonance are quite different in various models. In model I of Ref.~\cite{Wang:2017}, a second resonance $N(2000)5/2^+$ is found to provide considerable contributions at higher energies, while in the other two models, only one resonance, i.e., the $N(2060){5/2}^-$, is needed to describe the data. Different roles/contributions of nucleon resonances in these models clearly show how the resonance content and parameters depend on the way in which the background contributions are constructed for $\gamma p \to K^{*+}\Lambda$. These observations establish those obtained from the differential cross sections of Figs.~\ref{fig:dif1} and \ref{fig:dif2}.

\section{Summary and conclusions}  \label{sec:summary}

In this work, the high-precision differential cross-section data for the reaction $\gamma p\to  K^{*+} \Lambda$ are reanalyzed within a Regge-inspired effective Lagrangian approach, where the $t$-channel interactions are described by the Regge amplitudes or the interpolated Regge amplitudes instead of the Feynman amplitudes as in our previous work \cite{Wang:2017}. The purpose is to test how strongly the reaction mechanism, the extracted resonance contents and the associated resonance parameters depend on the way in which the $t$-channel meson-exchange amplitudes are constructed.

It is found that in both the Regge model and the interpolated Regge model, one only needs to introduce a single $N(2060){5/2}^-$ resonance in constructing the $s$-channel reaction amplitude in order to describe the cross-section data, which is quite different from the Feynman model of Ref.~\cite{Wang:2017}, where it was found that apart from $N(2060){5/2}^-$, the introduction of an additional resonance in the $s$ channel is indispensable to get an acceptable description of the data.

The reaction mechanisms are found to be highly model dependent. In the near-threshold region, especially at the center-of-mass energy $W=2041$ MeV, the angular distributions are dominated by the $K$ exchange in the Regge model, by the $N(2060){5/2}^-$ exchange in the interpolated Regge model, and by both the $K$ exchange and the $N(2060){5/2}^-$ exchange in the Feynman model of Ref.~\cite{Wang:2017}. At higher energies, the angular distributions are found to be dominated by the $K$ and $N(2060){5/2}^-$ exchanges in the Regge model, by the $N(2060){5/2}^-$ exchange in the interpolated Regge model, and by exchanging the $K$, the $N(2060){5/2}^-$, and other resonances in the Feynman model. The $K$ exchange, which plays a very significant role in almost the whole energy range considered in both the Regge model and the Feynman model, provides rather small contributions in the interpolated Regge model. In constrast, the $K^*$ exchange and the interaction current, which are negligible in both the Regge model and the Feynman model, offer considerable contributions in the interpolated Regge model.

The common feature of all these three models is that the $N(2060){5/2}^-$ resonance is needed and plays a very important role for the angular distributions. The fitted mass of this resonance is in a narrow range of about $2009$--$2043$ MeV and the fitted width varies in a broad range of about $57$--$213$ MeV in different models and various fits. This finding might be a hint that the $N(2060){5/2}^-$ is really needed and plays a significant role in the reaction $\gamma p\to K^{*+}\Lambda$, independently of how the $t$-channel amplitudes are constructed. The mass of this resonance can be basically determined by the present data, while the width cannot.

\begin{acknowledgments}
This work is partially supported by the National Natural Science Foundation of China under Grants No.~11475181 and No.~11635009, the Youth Innovation Promotion Association of Chinese Academy of Sciences under Grant No.~2015358, and the Key Research Program of Frontier Sciences of Chinese Academy of Sciences under Grant No.~Y7292610K1.
\end{acknowledgments}

\end{document}